\documentclass[aip,apl,reprint,superscriptaddress]{revtex4-1}
\usepackage{graphicx}
\usepackage{dcolumn}
\usepackage{bm}
\usepackage{amssymb}

\begin{document}


\title{Thermal Excitation of Broadband and Long-range Surface Waves on SiO$_2$ Submicron Films}


\author{Sergei Gluchko}
\affiliation{Laboratoire EM2C, CNRS, CentraleSup\'{e}lec, Universit\'{e} Paris-Saclay, Grande Voie des Vignes, 92295 Ch\^{a}tenay-Malabry CEDEX, France}

\author{Bruno Palpant}
\affiliation{Laboratoire de Photonique Quantique et Mol\'{e}culaire, CentraleSup\'{e}lec, ENS Paris-Saclay, CNRS, Universit\'{e} Paris-Saclay, Grande Voie des Vignes, 92290 Ch\^{a}tenay-Malabry, France}

\author{Sebastian Volz}
\affiliation{Laboratoire EM2C, CNRS, CentraleSup\'{e}lec, Universit\'{e} Paris-Saclay, Grande Voie des Vignes, 92295 Ch\^{a}tenay-Malabry CEDEX, France}
\affiliation{Laboratory for Integrated Micro-Mechatronic Systems, CNRS-IIS, The University of Tokyo, 4-6-1 Komaba, Meguro-ku, Tokyo}

\author{R\'{e}my Braive}
\affiliation{Centre de Nanosciences et de Nanotechnologies (C2N), CNRS, Universit\'{e} Paris-Saclay, Route de Nozay, F-91460 Marcoussis, France}
\affiliation{Universit\'{e} Paris Diderot, Sorbonne Paris Cit\'{e}, 75207 Paris CEDEX 13, France}

\author{Thomas Antoni}
\email[]{Corresponding author:thomas.antoni@centralesupelec.fr}
\affiliation{Laboratoire de Photonique Quantique et Mol\'{e}culaire, CentraleSup\'{e}lec, ENS Paris-Saclay, CNRS, Universit\'{e} Paris-Saclay, Grande Voie des Vignes, 92290 Ch\^{a}tenay-Malabry, France}



\date{\today}

\begin{abstract}
We detect thermally excited surfaces waves on a submicron SiO$_2$ layer, including Zenneck and guided modes in addition to Surface Phonon Polaritons. The measurements show the existence of these hybrid thermal-electromagnetic waves from near- (2.7~$\mu$m) to far- (11.2~$\mu$m) infrared. Their propagation distances reach values on the order of the millimeter, several orders of magnitude larger than on semi-infinite systems. These two features; spectral broadness and long range propagation, make these waves good candidates for near-field applications both in optics and thermics due to their dual nature.  
\end{abstract}


\maketitle

Thermal radiation through surface wave diffraction is usually only considered as the result of Surface Phonon Polaritons (SPhPs).
SPhPs are hybrid evanescent electromagnetic surface waves
generated by the phonon-photon coupling, at the interface of polar and dielectric
materials (such as SiO$_2$ and air) \cite{Joulain,1film,3film, NatureKim}.
The influence of SPhPs on the thermal performance of nanostructured materials has been studied intensively over
the last decade, providing an alternative channel of heat conduction when the objects are scaled down \cite{JoseThinFilm,ChenThinThermalPRB}. Due to this behaviour, they are essential for the improvement of the thermal stability in micro and nanoelectronics\cite{nanoelectr2,nanoelectr1,mikyung}, microscopy\cite{microscopyNature}, near-field thermophotovoltaics \cite{Gelais} and for
thermal radiation \cite{RousseauNature,nature,marquierPRB}. In addition, SPhPs provide coherent thermal radiation in mid-infrared \cite{nature,marquierPRB}.
This feature is now widely used to control thermal radiation but in a frequency range that is limited to the mid-infrared because it implies the coupling to transverse optical phonons \cite{Joulain,NatureBN}.
But this narrow spectrum (typically $8.6-9.3$~$\mu$m at a SiO$_2$-air interface) in addition to propagation lengths in the range of the wavelength decrease the field of use of SPhPs for many applications such as thermal transport at nanoscale, infrared nanophotonics and coherent thermal emission.

In this letter we demonstrate through experiment that coherent thermal emission, resulting from surface waves, can be extended spectrally. We also prove experimentaly that these surface waves have a long propagation range, when considering isolated submicron layers. Indeed, if the film is thinner than the penetration depth of the wave inside the material, the electromagnetic mode can be coupled on both its interfaces allowing for the long-range propagation of two other types of electromagnetic surface waves; Zenneck and subwavelength Transverse Magnetic (TM) guided modes \cite{yang}. The propagation length is increased as a consequence of the dramatic decrease in the overlap of the mode with the material, hence its absorption. For example, it is almost two orders of magnitude larger than the wavelength for a $1$~$\mu$m thick suspended SiO$_2$ membrane \cite{JoseThinFilm}. To prove those predictions, we fabricated a submicron glass layer and characterized its thermal emission by means of Fourier Transform InfraRed (FTIR) spectroscopy. Our experimental results are then compared with both Finite-Difference Time-Domain (FDTD) simulations and theoretical predictions.
 
\begin{figure}[t]
\includegraphics[width=230pt]{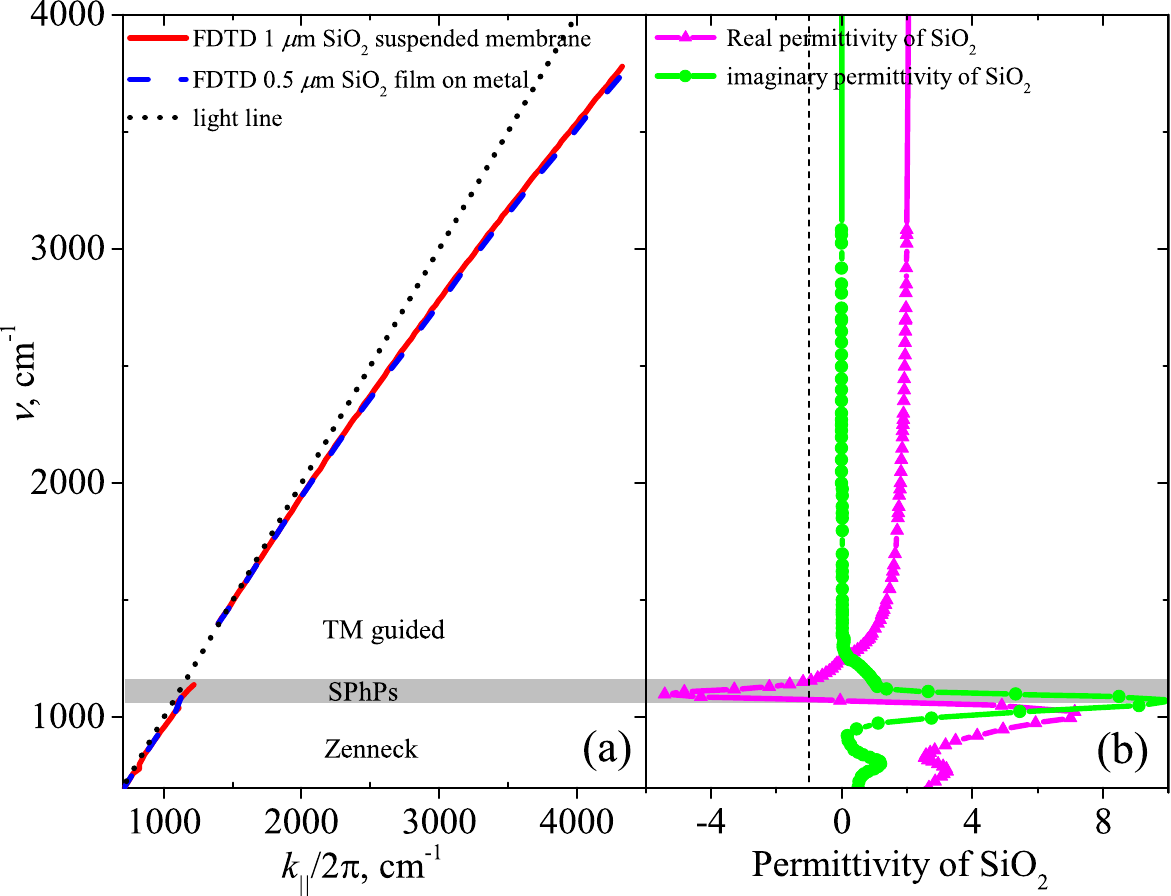}
\caption{\label{Fig.1} (a) Comparison of the FDTD-calculated dispersion relations of a 1~$\mu$m thick suspended SiO$_2$ membrane (red line) and of a 0.5~$\mu$m thick SiO$_2$ film deposited on aluminum (dashed blue line). (b) Real (magenta line) and imaginary (green line) parts of the relative permittivity of SiO$_2$. Grey region indicates the frequency range of SPhPs.}
\end{figure}

The fabrication process of a suspended submicron thick glass membrane is, however, a challenging task due to the poor selectivity of Si/SiO$_2$ etching. Nevertheless, as far as electromagnetism is concerned, a sample consisting of a twice thinner layer of SiO$_2$ deposited over a metallic film - the role of which is to optically isolate the dielectric thin layer from the substrate - is a strictly equivalent structure.
To prove this equivalence, Fig.~\ref{Fig.1}(a) reports the calculated dispersion curves of surface waves on a 1~$\mu$m thick suspended SiO$_2$ membrane and a $0.5$~$\mu$m thick SiO$_2$ thin film deposited on an aluminum layer obtained by means of FDTD simulations (MEEP code\cite{meep}). This is done by considering real and imaginary parts of glass dielectric function of SiO$_2$ shown in Fig.~\ref{Fig.1}(b) as extracted from experimental data \cite{palik}. 
The dispersion curves are superimposed, confirming the electromagnetic equivalence of the two systems. In addition, it can be seen that the curves lie beneath the light line over nearly the full spectrum indicating the evanescent behavior of the modes.   

\begin{figure}[b]
{\includegraphics[width=210pt]{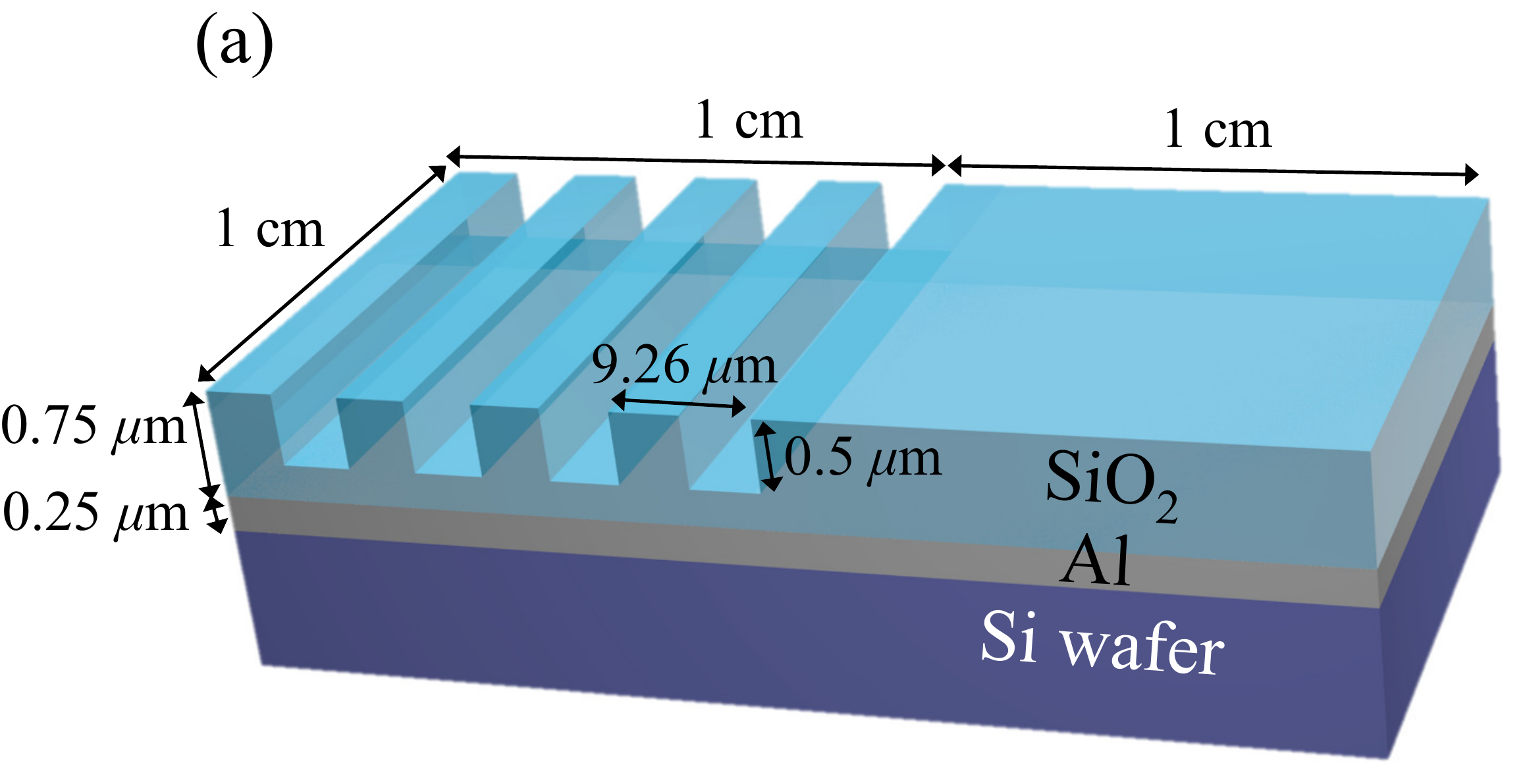}}

{\includegraphics[width=210pt]{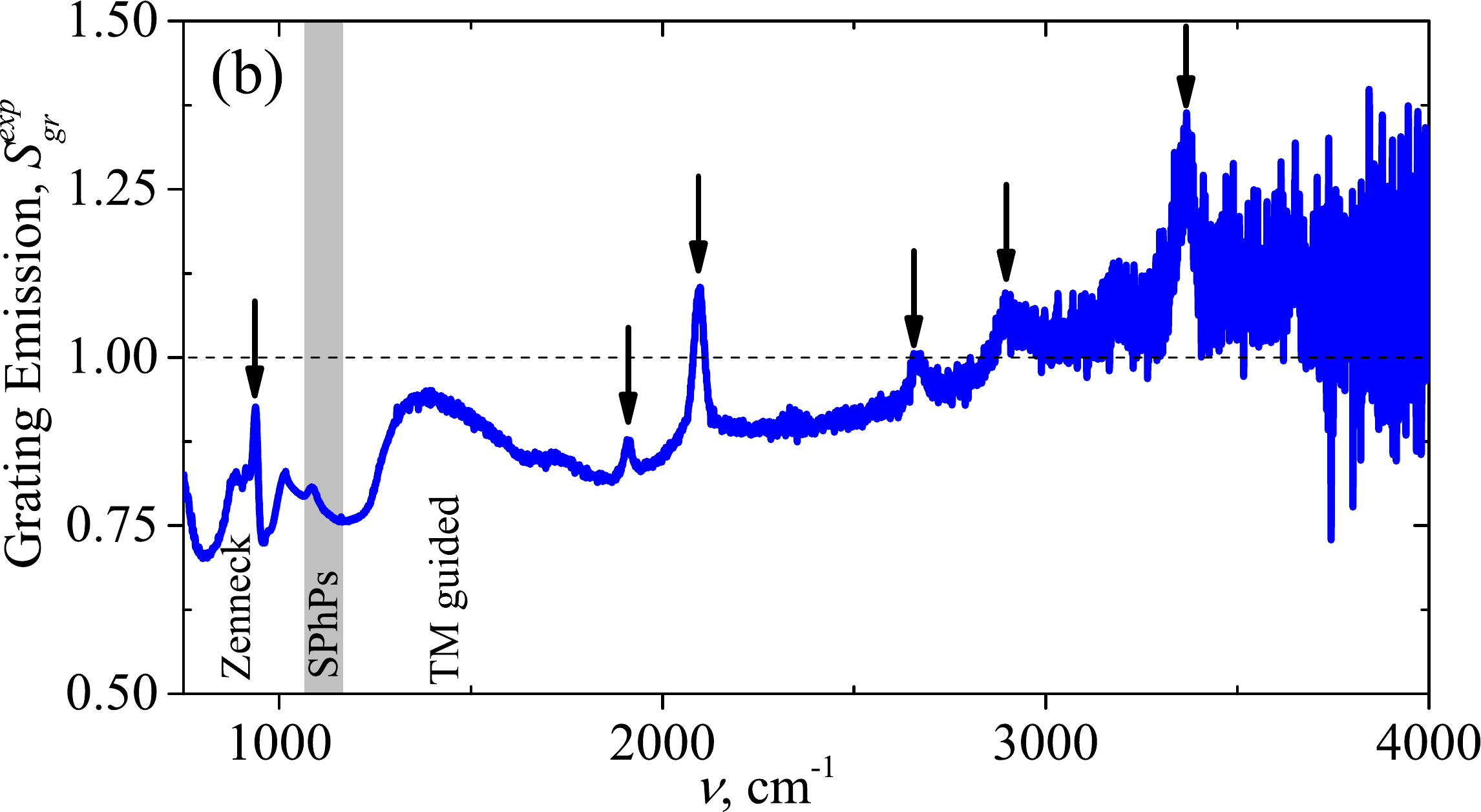}}
\caption{\label{Fig.3} Sample design (a) and its emission signal (b) at $T_s=673$~K and tilt angle $\alpha = 2.6^{\circ}$. Black arrows indicate the emission peaks due to the diffraction grating.}
\end{figure}

The sample is fabricated by sputtering deposition of an amorphous SiO$_2$ thin layer on an aluminum layer, itself grown on a polished Si wafer in order to optically mimic the suspended layer of SiO$_2$. The thicknesses of SiO$_2$ and aluminum layers are chosen to be $0.75$~$\mu$m and  $0.25$~$\mu$m, respectively. A diffraction grating of period $\Lambda = 9.26$~$\mu$m with a filling factor $0.5$ is etched by $0.5$~$\mu$m in the SiO$_2$ film using negative UV lithography and anisotropic reactive ion etching. In order to avoid any artificial broadening of the emission spectrum due to finite size effects, the grating has to be much larger than the propagation length of these surface modes. We then choose the lateral size of the grating to be 1~cm in both length and width (as shown on Fig.~\ref{Fig.3}(a)), that is one order of magnitude larger than the maximum value of the estimated propagation length.

We operate a FTIR spectrometer with a spectral resolution of $1$~$\text{cm}^{-1}$, which provides the required sensitivity in the working frequency range from $700$~$\text{cm}^{-1}$ to $4000$~$\text{cm}^{-1}$ ($14.3$ - $2.5$~$\mu$m). 
The sample is heated up to $673$~K with a heating stage and the emitted signal is collected for various tilt angles of the sample. KRS-5 holographic wire grid polarizer is used to examine sample emission for TE and TM polarizations providing excellent transmission in the working frequency range.
Each experimental spectrum $S^{exp}_{gr}$ is obtained by subtraction of the background radiation $S_{bg}$ and normalization by the emission obtained from a flat region of the sample, without grating $S_{fl}$, under the same conditions, as follows:
\begin{equation}
S^{exp}_{gr}(\nu, T_s,\alpha) = \frac{S_{gr}(\nu, T_s,\alpha)-S_{bg}(\nu)}{S_{fl}(\nu, T_s,\alpha)-S_{bg}(\nu)} \textnormal{,}
\label{Eq.1}
\end{equation}
where $T_s$ is the sample surface temperature and $\alpha$ is  the tilt angle. Note that this emission signal $S^{exp}_{gr}(\nu, T_s,\alpha)$ is not the same as emissivity as it also features the influence of the topography of the sample on its emission. The normalized spectrum value reduces to unity for any frequency where the grating has no impact on the sample emission. Fig.~\ref{Fig.3}(b) shows this normalized emission spectrum obtained from our sample for $T_s=673$~K. The sharp peaks, marked with black arrows, are the diffraction orders of the grating which is a clear signature of coherent thermal emission. The peaks without the arrows do not originate from the diffraction by the grating and they are observed due to the different SiO$_2$ effective thickness in the grating and flat regions of the sample.  

\begin{figure}[t]
{\includegraphics[width=110pt]{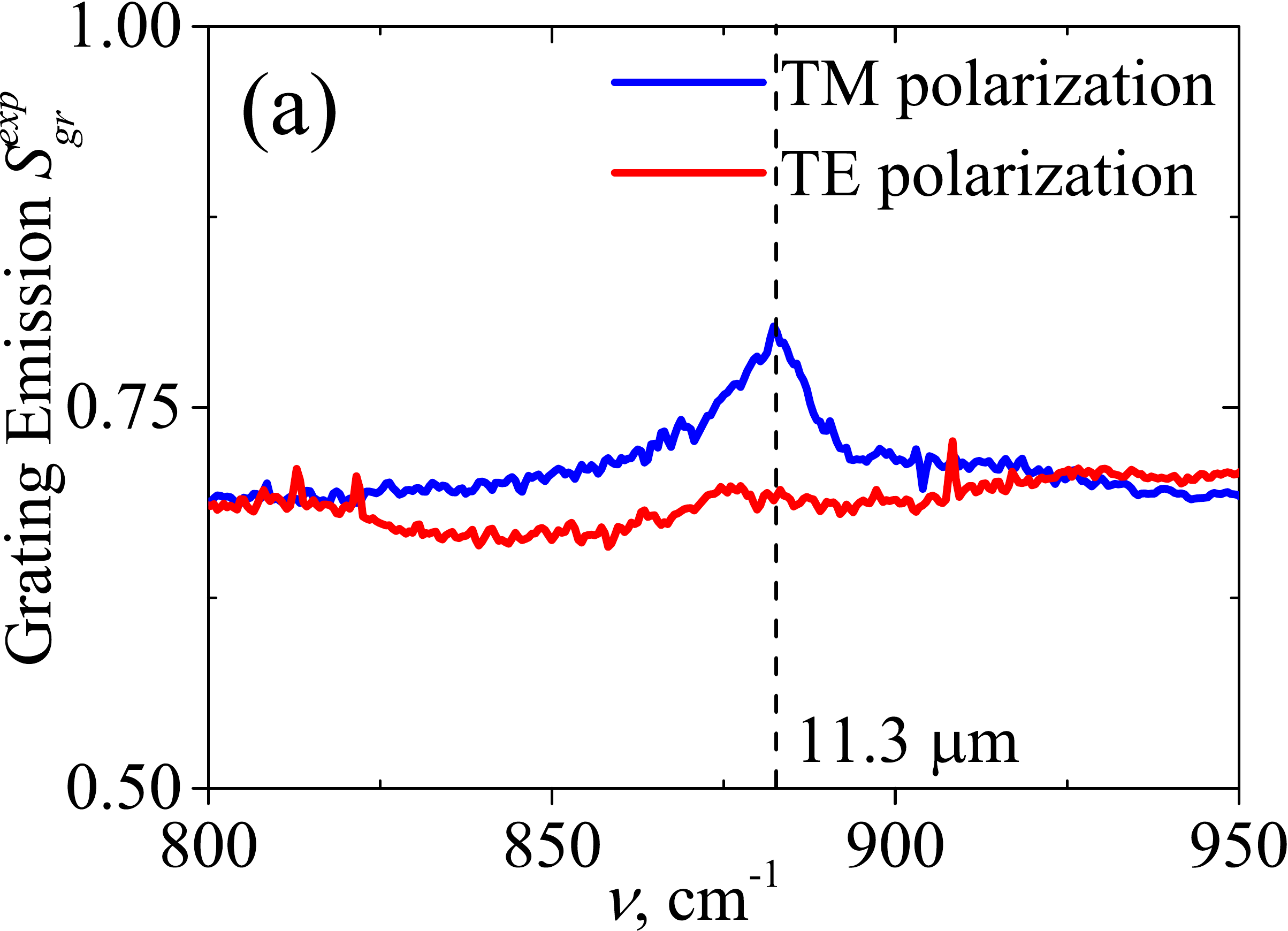}}%
{\includegraphics[width=110pt]{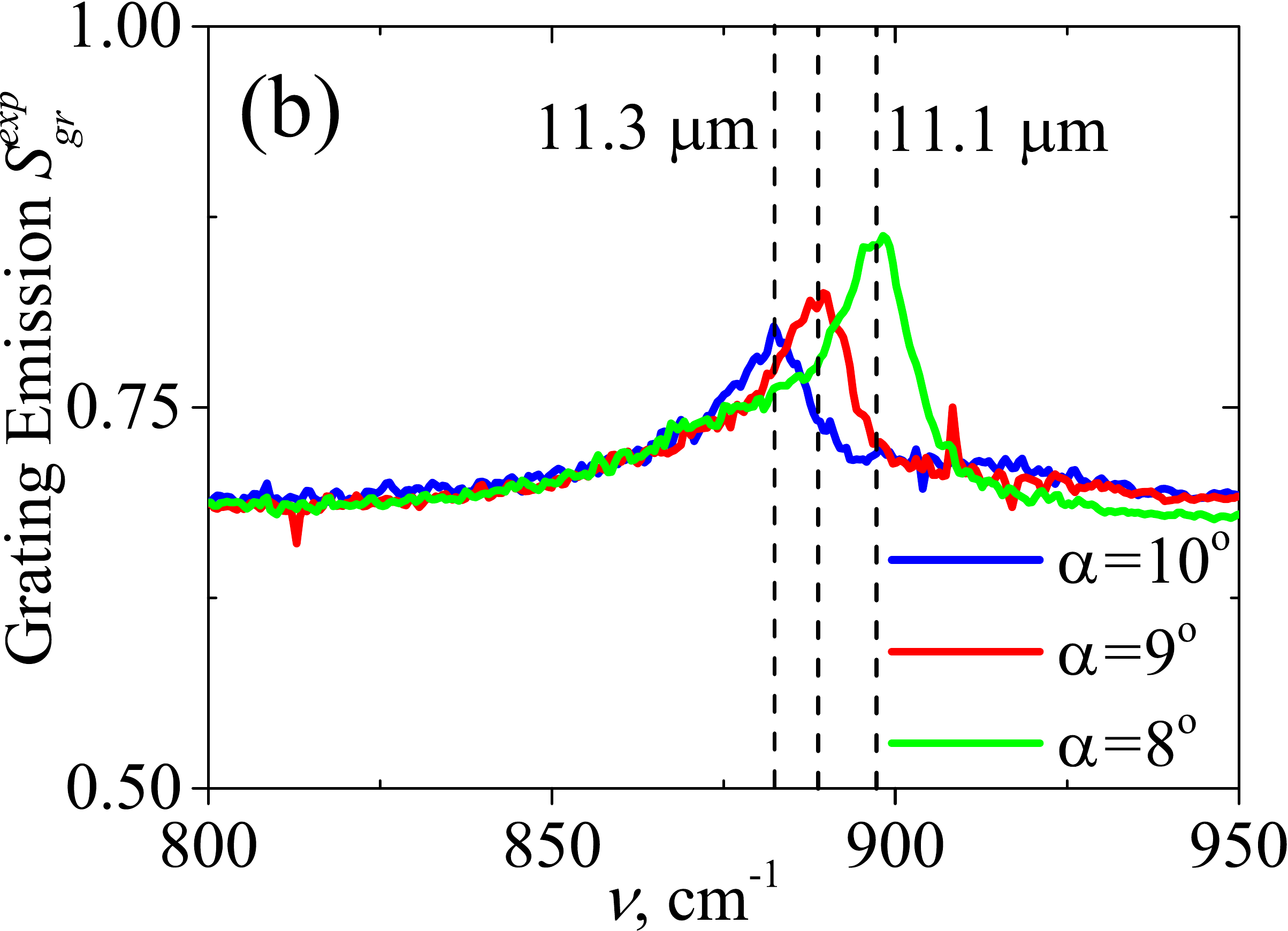}}
{\includegraphics[width=110pt]{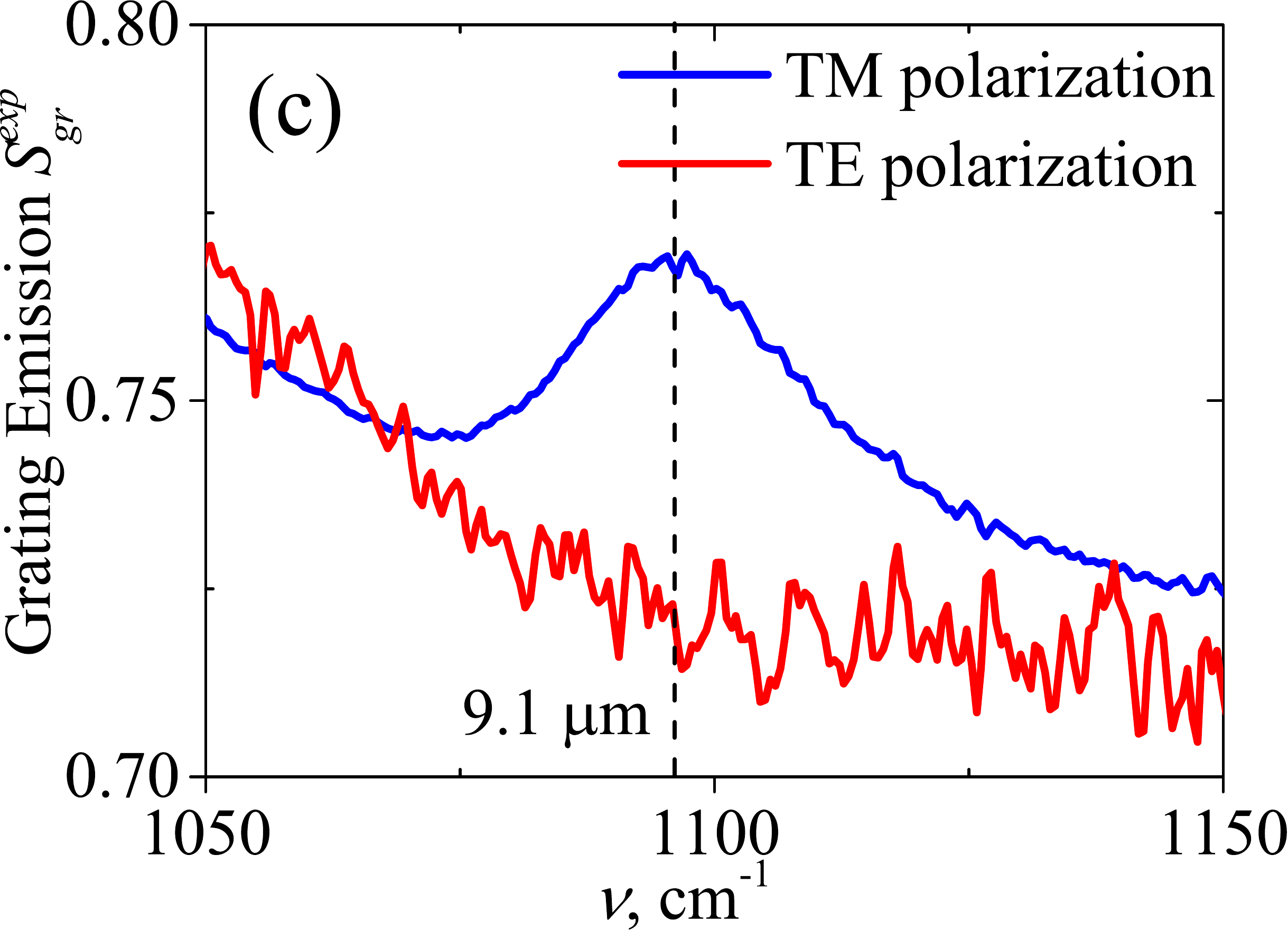}}
{\includegraphics[width=110pt]{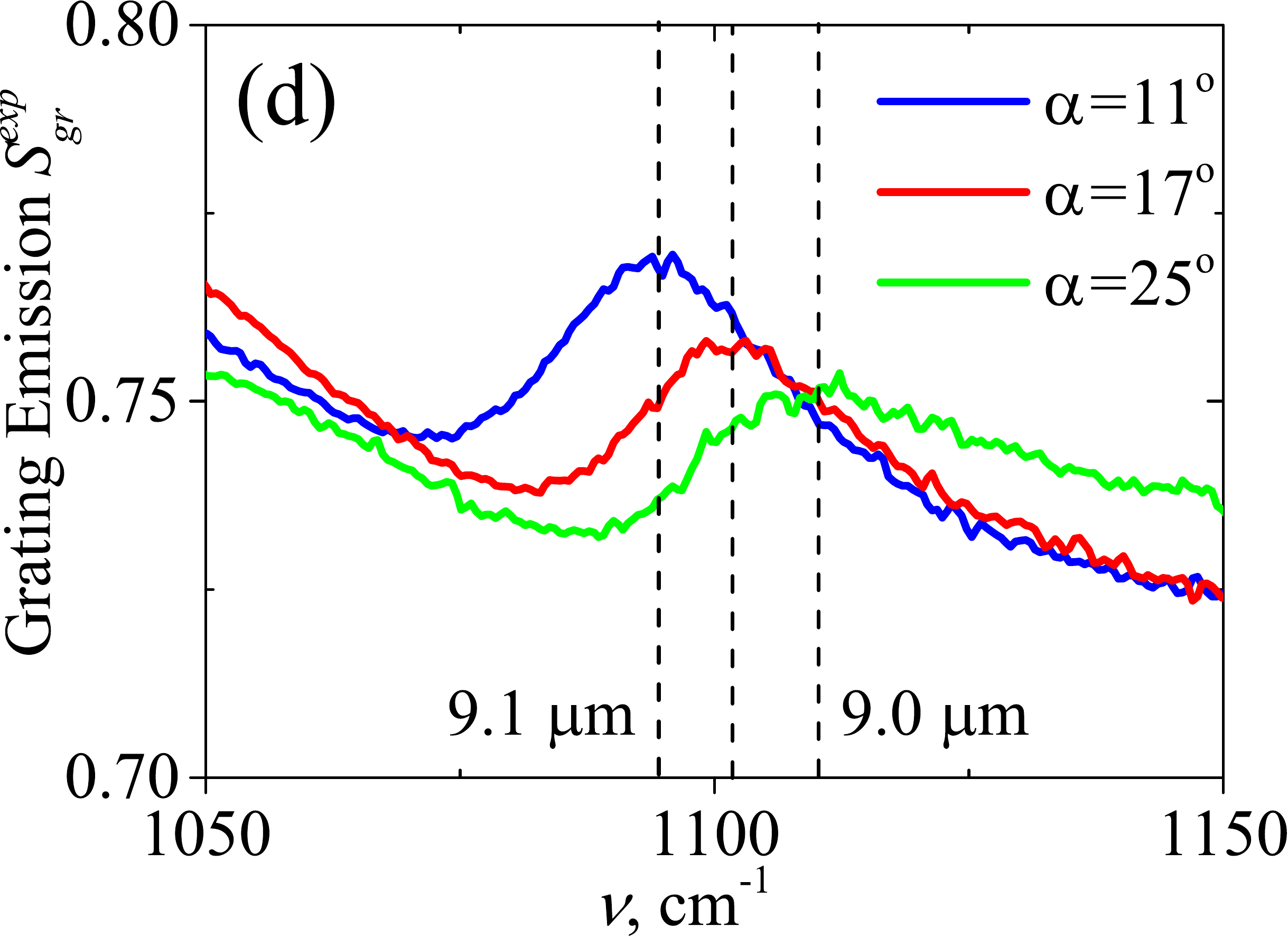}}
{\includegraphics[width=110pt]{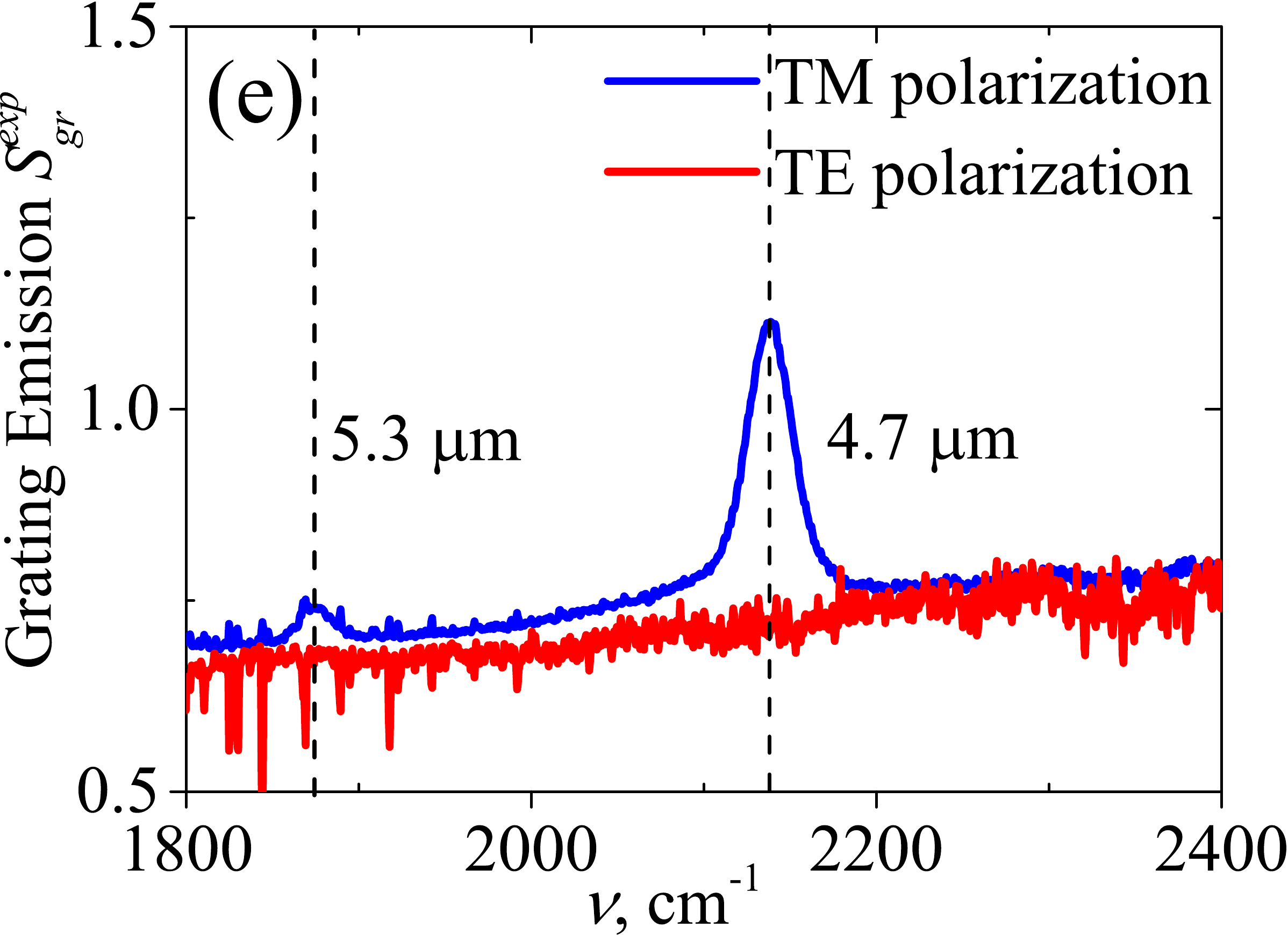}}
{\includegraphics[width=110pt]{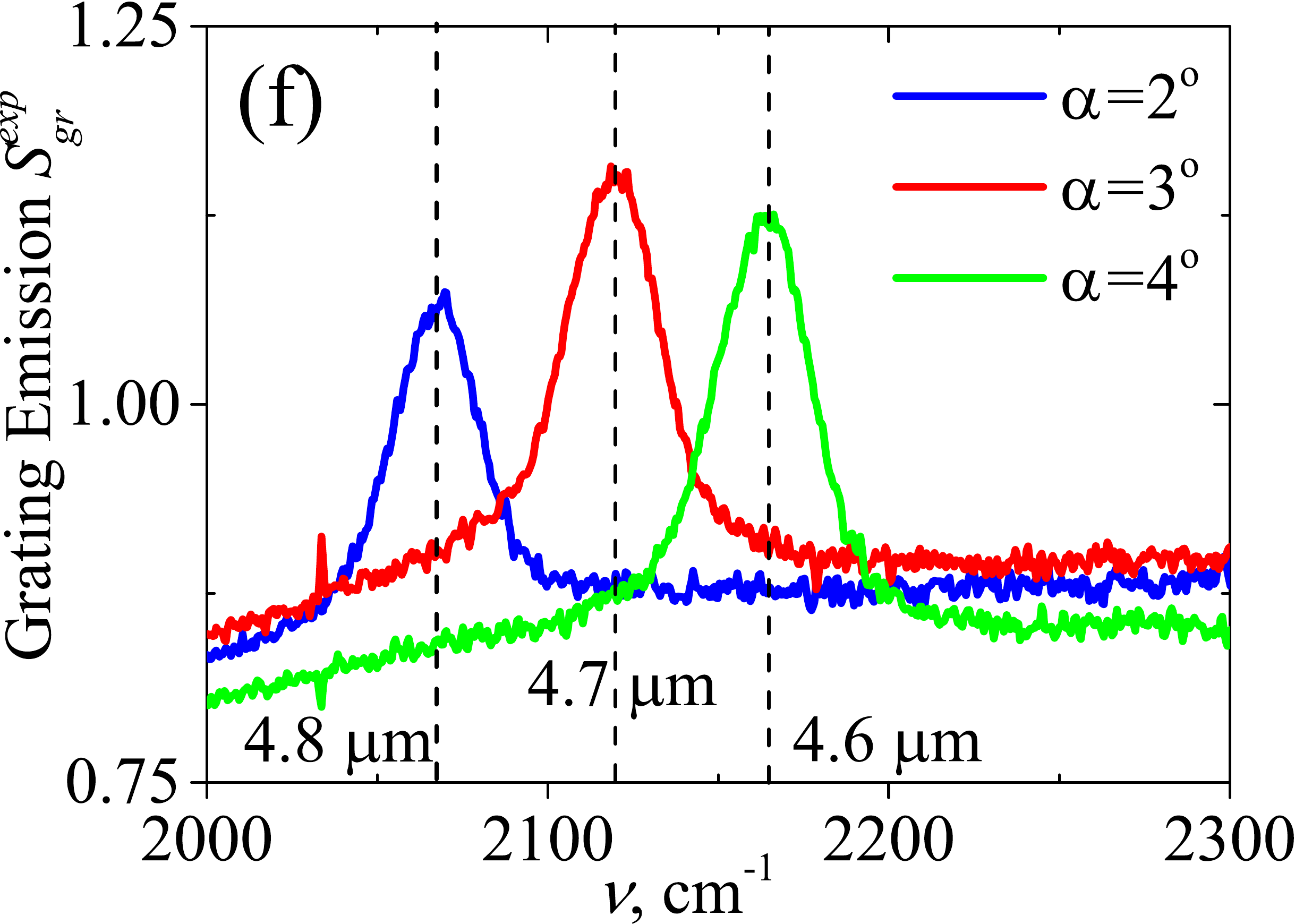}}
 \caption{\label{Fig.4}   
  Emission signal of the SiO$_2$ grating deposited on aluminum in TE and TM polarizations (left) and for different tilt angles (right). Figures (a) and (b): Zenneck region ($\epsilon_r > 0$, $\epsilon_i > 0$). Figures (c) and (d): Surface Phonon Polariton region ($\epsilon_r < -\epsilon_{\text{air}}$). Figures (e) and (f): subwavelength TM guided mode region ($\epsilon_r > 0$, $\epsilon_i \approx 0$).}
\end{figure}

We then examine the polarization of the grating emission since the diffraction peaks can only exist for transverse magnetic polarization if they originate from surface waves \cite{yang,book:SurfaceWavesBook}. 
Fig.~\ref{Fig.4}(a, c, e) show the grating emission peaks observed by collecting emission signal from the grating on SiO$_2$ thin film deposited on aluminum substrate heated up to $T=673$~K in three different frequency regions. In all these regions the emission peaks disappear for TE polarized signal as expected. 

These frequency regions indicate different surface modes according to the values of the relative permittivity of glass shown in Fig.~\ref{Fig.1}(b). These are Zenneck surface modes ($\epsilon_r > 0$, $\epsilon_i > 0$), Surface Phonon Polaritons ($\epsilon_r < -\epsilon_{\text{air}}$), and subwavelength TM guided modes ($\epsilon_r > 0$, $\epsilon_i \approx 0$) according to the classification of surface electromagnetic modes in thin dielectric films \cite{yang,book:SurfaceWavesBook}.
Detection of Zenneck surface modes in a single interface system for these frequencies is usually not possible due to their short propagation length \cite{yang} while a thin film supports long-range Zenneck modes allowing for their diffraction by the grating.
Fig.~\ref{Fig.4}(e) shows the existence of thermally excited subwavelength TM guided modes over a broad frequency range around $\nu=2000$~$\text{cm}^{-1}$ where the absorption is almost equal to zero.

\begin{figure}[b]
\includegraphics[width=220pt]{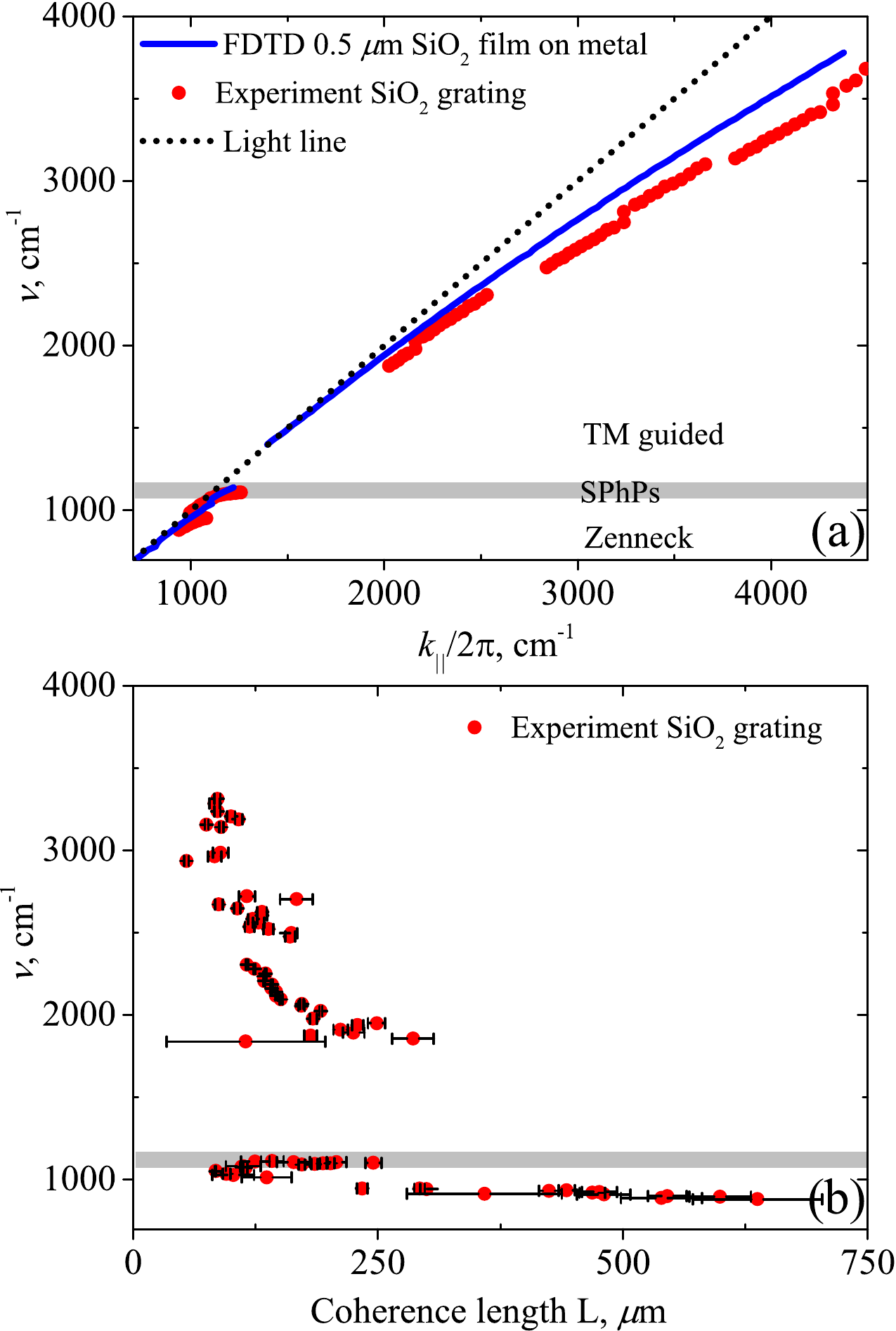}
\caption{\label{Fig.5} Dispersion relation (a) and coherence length (b) of the surface waves for the thin SiO$_2$ film deposited on aluminum obtained from FDTD direct computations (blue line) and experimental measurements (red circles).}
\end{figure}

Fig.~\ref{Fig.4}(b, d, f) demonstrate the frequency shift of the emission peaks when the tilt angle is varied. This feature can be used to reconstruct the dispersion relation of these surface modes by applying the grating equation:
\begin{equation}
\frac{\omega}{c}\sin{\alpha}=k_{\parallel}+m\frac{2\pi}{\Lambda}{,} \quad {m\in\mathbb{Z}}{,}
\label{Eq.2}
\end{equation}
where $k_{\parallel}$ is the in-plane wavevector and $m$ is the diffraction order. Fig.~\ref{Fig.5}(a) reports the dispersion relation that has been obtained by following this procedure. Experimental data lie beneath the light line indicating the evanescent nature of the surface waves and is in reasonably good agreement with the numerical predictions. The difference between experimental results and FDTD simulation data can be understood by assuming that the effective thickness of the SiO$_2$ film with the diffraction grating is not exactly 0.5~$\mu$m due to fabrication discrepancies and that the permittivities of the SiO$_2$ sample and of the FDTD computation differ. Note that we did not observe any diffraction peaks from $\nu=1300$~$\text{cm}^{-1}$ to $\nu=1900$~$\text{cm}^{-1}$ due to the presence of water vapor.

Lifetime estimation of these surface modes can be obtained by analyzing the width of the grating emission peaks. Considering the experimental dispersion to be photon-like, the coherence length can be deduced as follows:
\begin{equation}
L=\frac{1}{2\Delta\nu}{,}
\label{Eq.3}
\end{equation} 
where $\Delta\nu$ is the full wave number width at half maximum of the emission peak in $\text{cm}^{-1}$. Fig.~\ref{Fig.5}(b) shows that the typical coherence length is in the order of 100~$\mu$m, that is, ten times larger than the typical coherence length of these surface waves in a semi-infinite SiO$_2$-air interface. These values are in agreement with theoretical predictions for a 1~$\mu$m thick SiO$_2$ suspended film reported by Ordonez-Miranda \textit{et al.}\cite{JoseThinFilm}. The obtained length range is also ten times larger than the typical coherence length of the surface waves of a semi-infinite SiO$_2$-air interface \cite{JoseThinFilm, Joulain}. The coherence length $L$ reaches 700~$\mu$m for Zenneck modes which is almost two orders of magnitude larger than their wavelength. Such a large coherence length is achieved due to the fact that most of the electromagnetic energy is propagating in the air close to the interface of the dielectric rather than in the material, decreasing the absorbed power and enhancing the coherence length. Note, that we underestimate the coherence length of the modes since the grating obviously introduces radiative losses and that the values are expected to be even larger for smooth thin films.

In this work we have observed through experiment thermally excited surface waves at the surface of a thin SiO$_2$ film deposited on aluminum from $882$~$\text{cm}^{-1}$ to $3725$~$\text{cm}^{-1}$, whereas an interface between two semi-infinite materials only supports surface waves from $1072$~$\text{cm}^{-1}$ to $1156$~$\text{cm}^{-1}$. This spectral broadening is the result of Zenneck and subwavelength TM guided surface waves excitation from $882$~$\text{cm}^{-1}$ to $1072$~$\text{cm}^{-1}$ and from $1979$~$\text{cm}^{-1}$ to $3725$~$\text{cm}^{-1}$ in addition to Surface Phonon-Polaritons.
From their emission spectra, we were able to reconstruct their dispersion relation and to measure the coherence length of these waves. For Zenneck surface waves, it reaches almost 700~$\mu$m which is two orders of magnitude larger than their wavelength. We believe that because of their large coherence length as well as their existence in a very large spectrum, these surface waves can be considered for a wide range of applications in the infrared for both the thermal management of submicron structures and photonics engineering due to their dual nature.
Here we focused our experimental study on SiO$_2$ since it is a very common material in microelectronics but the same phenomena will exist for any dielectric material supporting resonant surface waves.

We wish to thank Jose Ordonez-Miranda, St\'{e}phane Collin, Laurent Tranchant and Mikyung Lim for their fruitful discussions. We also want to acknowledge Thuy-Anh Nguyen, Paul Debue and Romaric de L\'{e}pinau for their contribution to the early development of the FDTD code. This work was supported by Renatech project ``Phonons Polaritons de Surface large bande''.

\bibliography{manuscript}

\end{document}